RESEARCH PAPER

# Density-dependent and independent mechanisms jointly reduce species performance under nitrogen enrichment

David Sampson Issaka[1], Or Gross[1], Itunuoluwa Ayilara[1], Talia Schabes[1], Niv DeMalach[1*],

[1]Institute of Plant Sciences and Genetics in Agriculture, The Hebrew University of Jerusalem, Rehovot, Israel

* Corresponding author: Niv.demalach@mail.huji.ac.il

**ABSTRACT**

Nitrogen (N) deposition is a primary driver of species loss in plant communities globally. However, the mechanisms by which high N availability causes species loss remain unclear. Many hypotheses for species loss with increasing N availability highlight density-dependent mechanisms, i.e., changes in species interactions. However, an alternative set of hypotheses highlights density-independent detrimental effects of nitrogen (e.g., N toxicity). We tested the role of density-dependent and density-independent mechanisms in reducing species performance. For this aim, we used 120 experimental plant communities (mesocosms) comprised of annual species growing together in containers under four fertilization treatments: (1) no nutrient addition (control), (2) all nutrients except N (P, K, and micronutrients), (3) Low N (3gN $m^{-2}$) + other nutrients, and (4) high N (15gN $m^{-2}$) + other nutrients. Each fertilization treatment included two sowing densities to differentiate between the effects of competition (N × density interactions) and other detrimental effects of N. We focused on three performance attributes: the probability of reaching the reproduction period, biomass growth, and population growth. We found that individual biomass and population growth rates decreased with increasing sowing density in all nutrient treatments, implying that species interactions were predominantly negative. The common grass (*Avena barbata*) had a higher biomass and population growth under N enrichment, regardless of sowing density. In contrast, the legume (*Trifolium purpureum*) showed a density-independent reduction in biomass growth with increasing N. Lastly, the small forb (*Silene palaestina*) showed a density-dependent reduction in population growth, i.e., the decline occurred only under high density. Our results demonstrate that density-dependent and density-independent mechanisms operate simultaneously to reduce species performance under high N availability. Yet, their relative importance varies among species and life stages.

## ACKNOWLEDGEMENTS

The study was supported by the Israel Science Foundation Grants no. 2403/22 and 672/22 . DSI was supported by the Pears Foundation. TS was supported by the School of Environmental studies of the Hebrew University. We thank Joanna Ouaknine for her help in the fieldwork. Tyler Poppenwimer, Nir Band, and two anonymous reviewers have provided constructive on earlier versions of the manuscript.

## AUTHOR CONTRIBUTIONS

ND and OG designed the experiment. DSI, IA, TS, and OG performed the experiment and collected the data. DSI analyzed the data. DSI and ND wrote the first draft, and all other authors substantially contributed to the writing of the manuscript.

## DATA AVAILABILITY

The data and codes for this manuscript are available in Figshare,
https://figshare.com/account/home#/collections/6183493

## 1. INTRODUCTION

Human activities have elevated Nitrogen (N) deposition in various ecosystems (Simkin et al. 2016, Ackerman et al. 2019, Borer and Stevens 2022). This increase in N availability is associated with a decline in species diversity, especially in grassland communities (Stevens et al. 2004, Harpole et al. 2016, Midolo et al. 2019, Band et al. 2022, Eskelinen et al. 2022). Moreover, the magnitude of diversity loss following N addition often increases with spatial scale (Seabloom et al. 2021, but see Lan et al. 2015) and can be irreversible (Isbell et al. 2013). Currently, N deposition is considered the third greatest driver of biodiversity loss worldwide after climate and land-use changes (Sala et al. 2000, Bobbink et al. 2010, Borer and Stevens 2022)

So far, numerous N addition experiments have been conducted worldwide, showing that elevated N levels lead to species loss and decline of forbs and legumes (Bobbink et al. 2010, Soons et al. 2017, Midolo et al. 2019, Tognetti et al. 2021, Band et al. 2022). Two mechanisms can drive such species loss within the framework of the classical niche theory (Hutchinson 1957). First, high N levels can fall outside the *fundamental niche* of some species, i.e., they cannot persist regardless of the density of other species. Alternatively, elevated N levels can fall outside the *realized niche* of some species (extinction due to a competitive exclusion). Therefore, a better understanding of the effects of elevated N requires separating density-dependent effects (competition) vs. density-independent detrimental effects (e.g., Chase & Leibold, 2003, Thompson et al., 2020).

Many studies investigating community response to N enrichments have highlighted density-dependent mechanisms, i.e., changes in competitive interactions with increasing N availability. First, an increase in standing biomass enhances light competition, thereby reducing short species' competitive ability (Eek and Zobel 2001, Hautier et al. 2009, Lamb et al. 2009, Borer et al. 2014a, DeMalach et al. 2016, Eskelinen et al. 2022). Nonetheless, N addition can also intensify belowground competition (Rajaniemi 2003a, Dickson and Foster 2011). Furthermore, eliminating the N limitation can decrease the potential for niche partitioning among species (niche dimensionality, sensu Harpole et al., 2016). Additionally, N addition can increase litter production, thereby reducing seedlings' establishment (Tilman, 1993, Foster & Gross, 1998). While there is no consensus on the relative importance of the above competitive exclusion mechanisms (Rajaniemi 2003b, Dickson and Foster 2011, DeMalach and Kadmon 2017, Harpole et al. 2017) all these mechanisms have one shared attribute of being density-dependent.

An alternative set of explanations does not invoke resource competition. Instead, it highlights the detrimental effects of N, which do not necessarily depend on density ("the nitrogen detriment hypotheses" sensu Band et al., 2022). For example, N addition can lead to ammonium toxicity (Britto and Kronzucker 2002), soil acidification (Crawley et al. 2005) which prevents the acquisition of nutrients (Tian et al. 2022), increased susceptibility to stress agents (Bobbink *et al.*, 2010, Stevens *et al.*, 2010), and an alternation of the soil microbiome (Farrer and Suding 2016, Huang et al. 2021). Again, these specific mechanisms are difficult to disentangle and are highly variable among systems. For example, differences in community sensitivity to N-induced acidification depend on soil buffering capacities (Clark et al., 2007). Still, all these mechanisms can potentially reduce performance independent of species interactions.

Recently, Band et al., (2022) conducted a global meta-analysis of ~600 experiments using biomass as a proxy for density-dependent effects. They found that the biomass-mediated effect of N leads to an average decline of ~2% in diversity, while the biomass-independent effect causes a more substantial diversity decline of ~18%. These findings are surprising given that many competition experiments have established a causal link between nitrogen addition, competition, and lower species diversity(Gurevitch et al. 1990, Lamb et al. 2009, DeMalach et al. 2017b).

One possible explanation for the results of Band et al. (2022) is that biomass is not a good proxy for density-dependent processes. Many studies have shown that biomass is affected by multiple factors (resources, nutrients, herbivores, species composition, litter production) and that the causality of the biomass-richness relationship is bidirectional (Grace et al. 2016). Alternatively, it is possible that even within the same community, some species experience reduced performance due to density-independent processes while other species experience reduced performance due to competition. Surprisingly, despite the numerous N-addition experiments, we are unaware of any experiment that separated the density-dependent and independent effects of N enrichment on the performance of different species.

Here, we manipulated sowing density as the most direct approach to disentangle density-dependent and independent processes. This simple manipulation has been widely used for monocultures but rarely for communities (but see Goldberg & Estabrook, 1998, Rajaniemi, Turkington, & Goldberg, 2009). We used this approach to investigate the effects of N availability on species performance within an experimental community that included a small forb (*Silene palaestina*), a medium size legume (*Trifolium purpureum*), and a tall competitive grass (*Avena barbata*).

We expected that the tall grass would benefit from N addition because tall grasses are often strong competitors for light (Suding et al. 2005, Vojtech et al. 2007, Gough et al. 2012, DeMalach et al. 2017b) and are insensitive to the detrimental effects of high N availability (Tian et al. 2022). In contrast, since the abundance of forbs and legumes often decreases with N availability (Suding et al. 2005, Tognetti et al. 2021), we hypothesized that N addition would increase their sensitivity to competition (competitive response sensu Miller & Werner, 1987). Additionally, we expected the legume and forb species to experience density-independent detrimental effects of N addition (Band et al. 2022). Still, we predicted that the density-independent effect would be minor compared to competition within the scope of our short-term experiment (see also Eskelinen et al. 2022) because toxic effects in the soil often accumulate over time (Tian et al. 2020).

## 2. MATERIALS AND METHODS

### 2.2 Experimental design

The experiment was conducted at the experimental station of the Hebrew University of Jerusalem in Rehovot, Israel (31°54'18.8 "N, 34°48'16.9 "E). The station is characterized by a Mediterranean climate, with a mild, rainy winter and a hot, dry summer (annual rainfall of ~560mm yr$^{-1}$).

The experiment mimicked the annual plant communities that grow on Hamra soils of Israel's coastal region. These communities grow during winter, bloom in spring, and dry in the late spring. Soil type is a primary driver of species composition in these communities. Communities growing on a sandy-type Hamra are characterized by high diversity and abundance of forbs and legumes. In contrast, loamy-type Hamra (with higher nutrients and water availability) has a lower diversity and high dominance of tall grasses.

The experimental plant communities were sown in January 2021 in large plastic containers (dimensions of ~1m × 1m × 1m) with holes for water drainage (Figure 1). The experiment included 120 experimental communities (3 soil types × 4 nutrient treatments × 2 sowing densities × 5 replications) randomly located in an area of ~half a hectare. We filled the containers with soils (taken from at least 1m depth to avoid seed bank) from three nearby locations aiming for a gradient of clay content (from sandy to loamy soil). However, two sources had almost identically low clay content (and water-holding capacity). Thus, to increase statistical power, these two sources were pooled into a single category (hereafter sandy soil ~7.5% clay content) and contrasted with the other source, red soil (hereafter loamy soil ~ 15% clay content).

We chose four model species that were collected in Israel's coastal plain (by a commercial company, 'Hila Pirchei Bar Ltd'): *Silene palaestina* (Caryophyllaceae, seed mass of 0.095mg), *Trifolium purpureum* (Fabaceae, 1.47mg), *Avena barbata* (Poaceae, 15.12mg), and *Lupinus palaestinus* (Fabaceae, 257.3mg). The species were chosen based on the following criteria: (i) being common in Hamra communities such that a sufficient amount of seeds was available (1000g per species). (ii) Representing different functional groups (forbs, legumes, and grasses) and a wide range of seed masses.

We fertilized the plots a few days after sowing in January 2021. The four fertilization treatments included: (i) no nutrient addition, (ii) the addition of all nutrients except N (P, K, and micronutrients following NutNet protocol, Borer *et al*., 2014), (iii) Low N (3gN m$^{-2}$, applied as urea) together with all other nutrients, and (iv) high N (15gN m$^{-2}$) with all other nutrients. Since we focused on N, we removed the potential limitation of all other nutrients by adding them to all plots and viewed treatment ii as another 'control'. However, to quantify the potential limitations of other nutrients, we also included a standard control with no addition. Hereafter, we refer to the no-nutrient treatment as the control to avoid confusion.

Under each nutrient treatment, the plant communities (all four species) were sown in two densities, low (3g m$^{-2}$) and high (16g m$^{-2}$), separated equally across the four species (in terms of weight). Sowing species in equal weight (rather than equal seed number), implies that the number of sown seeds is higher for small-seeded species (and vice versa). The rationale of this choice is that small-seeded species have a lower emergence probability (Table S1) and lower survival probability of seedlings (Tables S2, S3).

We chose 16g m$^{-2}$ as the high sowing density following previous studies of annual plant communities using this amount as the highest end of a density gradient (Godoy et al. 2014, Pérez-Ramos et al. 2019). The low-density level was chosen to reduce competitive interactions (although we did not assume that interactions are entirely eliminated). At the same time, a major requirement of our experiment was that even under low density, there would be enough individuals to calculate species' performance. Thus, we have sown only four species (having more species would reduce each species' density, i.e., separating the 3g among more species). In the experiment, the densities were sufficient for all species except *Lupinus*. *Lupinus* was missing in many (low-density) plots and excluded from the analysis.

Throughout the growing season, we weeded all species that were not part of the experiment. Additionally, the experiment was irrigated weekly during the spring (March-April). Otherwise, the containers would dry faster than the surrounding vegetation because of their high drainage (DeMalach, Ron, & Kadmon, 2019).

## 2.3 Sampling

We estimated population density by sampling seedlings during February 2021 and February 2022. Sampling was conducted in four 20cm × 20cm quadrates in each plot in the first year. However, due to logistic constraints, we sampled in two quadrates only for *Silene* and *Trifolium* and in four quadrates for *Avena*. During the second-year sampling, we avoided spots where biomass was collected in the first year.

To quantify individual biomass, we collected biomass samples from two 20cm × 20cm quadrates in the center of each plot (at the end of the first growing season, May 2021). First, the biomass was oven-dried at $60^0$C for 48hrs and sorted into species. Then, we counted the number of individuals of each species and weighed their biomass. In the analysis, all samples within a plot were aggregated (by taking the mean). Lastly, density units were converted from individuals per sample into individuals per square meter (i.e., multiplied by 25, the ratio between a quadrate area and a square meter).

## 2.4 Statistical analyses

All statistical analyses were conducted with R (version 4.1.1). We tested the effects of the treatments on the probability of reaching reproduction (the ratio between the number of sown seeds and the number of adults in the first year), biomass growth (the ratio between adult and seed masses), and population growth (the ratio between seedling densities in two consecutive years).

For each species in each soil type, we built the following linear regression model using dummy variables to describe the four nutrient treatments and the two sowing density levels:

$$Y = \beta_0 + \beta_1 Density + \beta_2 Micro + \beta_3 N3 + \beta_4 N15 + \beta_5 (Density \times N15)_{ij} + \varepsilon, \quad \varepsilon \sim N(0, \sigma_\varepsilon^2)$$

Here, $Y$ is a performance attribute (reproduction probability, biomass growth, or population growth rate). *Density* represents the sowing density (low[0] vs. high[1]), and *Micro* indicates whether or not PK and micronutrients were added. The variables *N3* and *N15* represent the addition of low and high levels of N. Using dummy variables rather than a continuous variable of N level was needed to avoid assuming a linear response to N addition.

The model includes an interaction term between sowing density and high levels of N (interactions with other nutrient treatments were insignificant and therefore removed to increase the statistical power and avoid overfitting). Notably, a significant interaction between N addition and sowing density is interpreted as evidence of a density-dependent response to N (i.e., N addition modifies competitive interactions). In contrast, when N has a significant effect, but there were no interactions, we interpreted it as evidence that the response to N addition is primarily density-independent.

To ensure normally distributed residuals in the linear regression ($\varepsilon \sim N(0, \sigma_\varepsilon^2)$), biomass and population growth were log$_e$-transformed while reproduction probability was logit-transformed ($logit(p) = \log_e(\frac{p}{1-p})$). To avoid zeros in the logit transformation (leading to undefined values), we added 0.01 to all observations. Additionally, for *Avena*, there were a few cases where reproduction probability was equal to or greater than one (because the density of the whole plot was extrapolated from quadrates and spatial distribution was non-homogenous), and those values were converted to 0.99 (logit of one is also undefined).

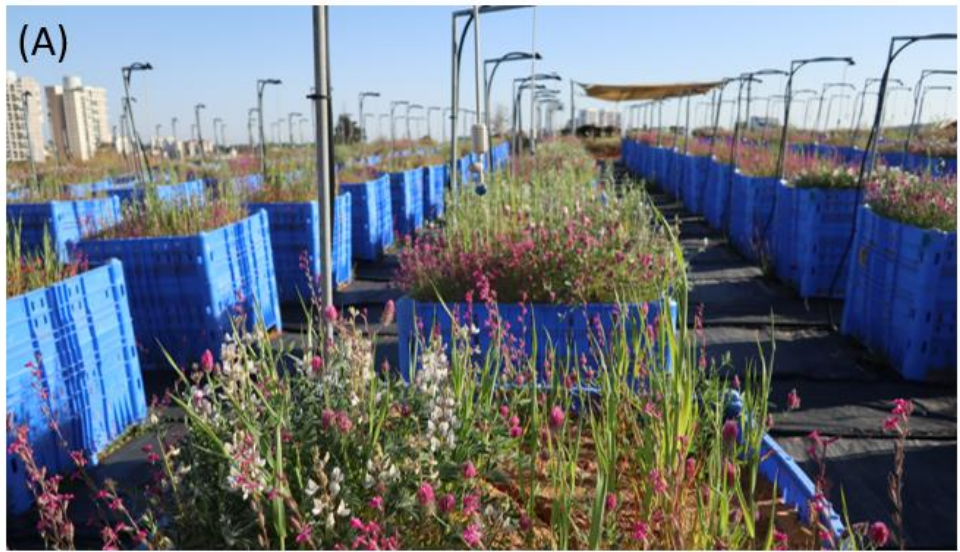
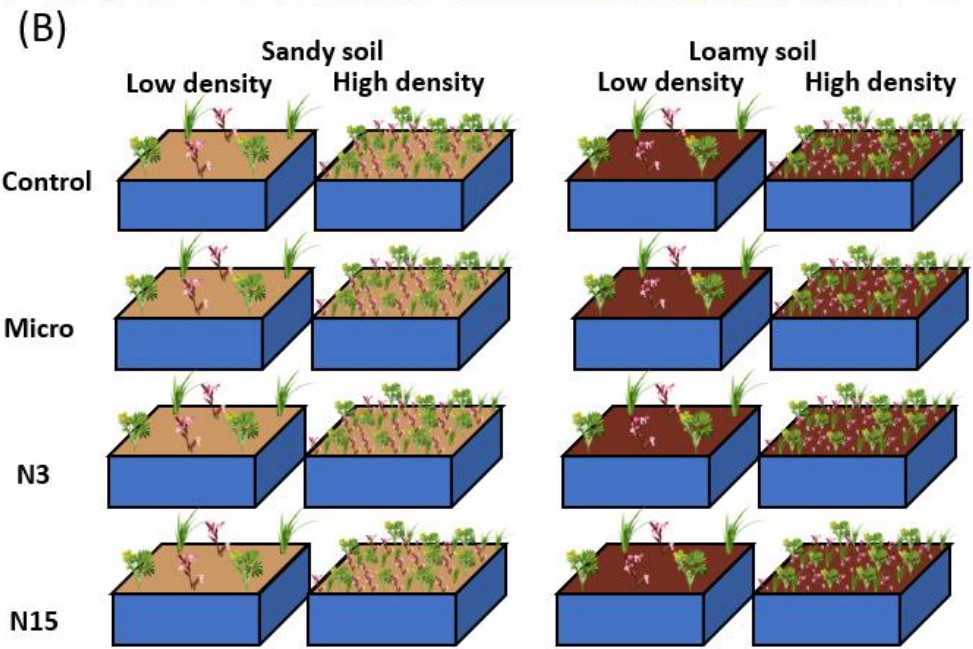

**Figure 1. (A) a photo of the experiment (B) a scheme of the experimental design. Two sowing densities (low vs. high) were applied to quantify both density-independent and density-dependent performance. Besides density manipulation, the experiment included a factorial combination of two soil types (sandy vs. loamy) and four fertilization treatments: (1) no nutrient addition (control), (2) the addition of all nutrients (P, K, and micronutrients) except N (Micro), (3) Low N together with all other nutrients (N3), and (4) high N with all other nutrients (N15).**

## 3. RESULTS

Reproduction probability, the probability of the seed reaching the reproduction period, was low in the small-seeded *Silene*, intermediate in *Trifolium*, and highest for the large-seeded *Avena* (Fig. 2, Tables S2 and S3). The effect of sowing density was negative for *Silene* ($P_{sandy} < 0.001$, $P_{loamy} = 0.004$), positive for *Trifolium* in the sandy soil ($P < 0.001$), and insignificant for *Avena*. The effects of N on reproduction probability were insignificant except for a marginally significant positive effect on *Silene* in the sandy soil ($P_{N15} = 0.065$). The addition of other nutrients reduced the reproduction probability of *Trifolium* in the sandy soil ($P = 0.003$) and increased it for *Silene* in the loamy soil ($P = 0.04$).

Biomass growth, the ratio between adult and seed biomass, was highest in the small-seeded *Silene*, intermediate in *Trifolium*, and lowest for the large-seeded *Avena* (Fig 3). The effects of increasing sowing density on biomass growth were strongly negative for all the species indicating the dominance of competitive interactions in the system (Table S4, S5). High levels of nitrogen addition increased the biomass growth of *Silene* ($P_{loamy} = 0.04$) and *Avena* ($P_{loamy} = 0.06$). However, the other nutrients did not affect biomass growth except for a positive effect on *Trifolium* ($P_{loamy} = 0.004$). Importantly, the biomass growth of *Trifolium* was reduced in the sandy soil under the high nitrogen addition ($P_{sandy} = 0.058$) and in loamy soils under low and high N addition ($P_{loamy, N3} = 0.038$, $P_{loamy, N15} = 0.019$). Strikingly, in low-density plots where all nutrients were added, *Trifolium*'s biomass growth was reduced from ~510 to ~210 and ~140 under low and high nitrogen levels, respectively. Moreover, we found no interactions between N and density, indicating that the detrimental effects of N were density-independent.

Increasing sowing density reduced the population growth (the ratio between seedling densities in the two years) of all species, demonstrating that competition strongly affects population dynamics (Fig 4, Tables S6, S7). Notably, *Silene* had strong interactions between growth rate and density ($P_{loamy, N15*Density} = 0.0019$). Under high nitrogen, its population growth rate was reduced from 2 (population increase) under low density to 0.4 (population decline) under high density. In contrast, the population growth of *Trifolium* was unaffected by N (but was reduced in the loamy soil). Lastly, for *Avena*, the population growth increased under high N levels in the sandy soils regardless of sowing density ($P_{sandy} = 0.012$). Additionally, we found no effect of adding other nutrients on species' population growth, except for a marginal increase for *Trifolium* ($P_{loamy} = 0.06$).

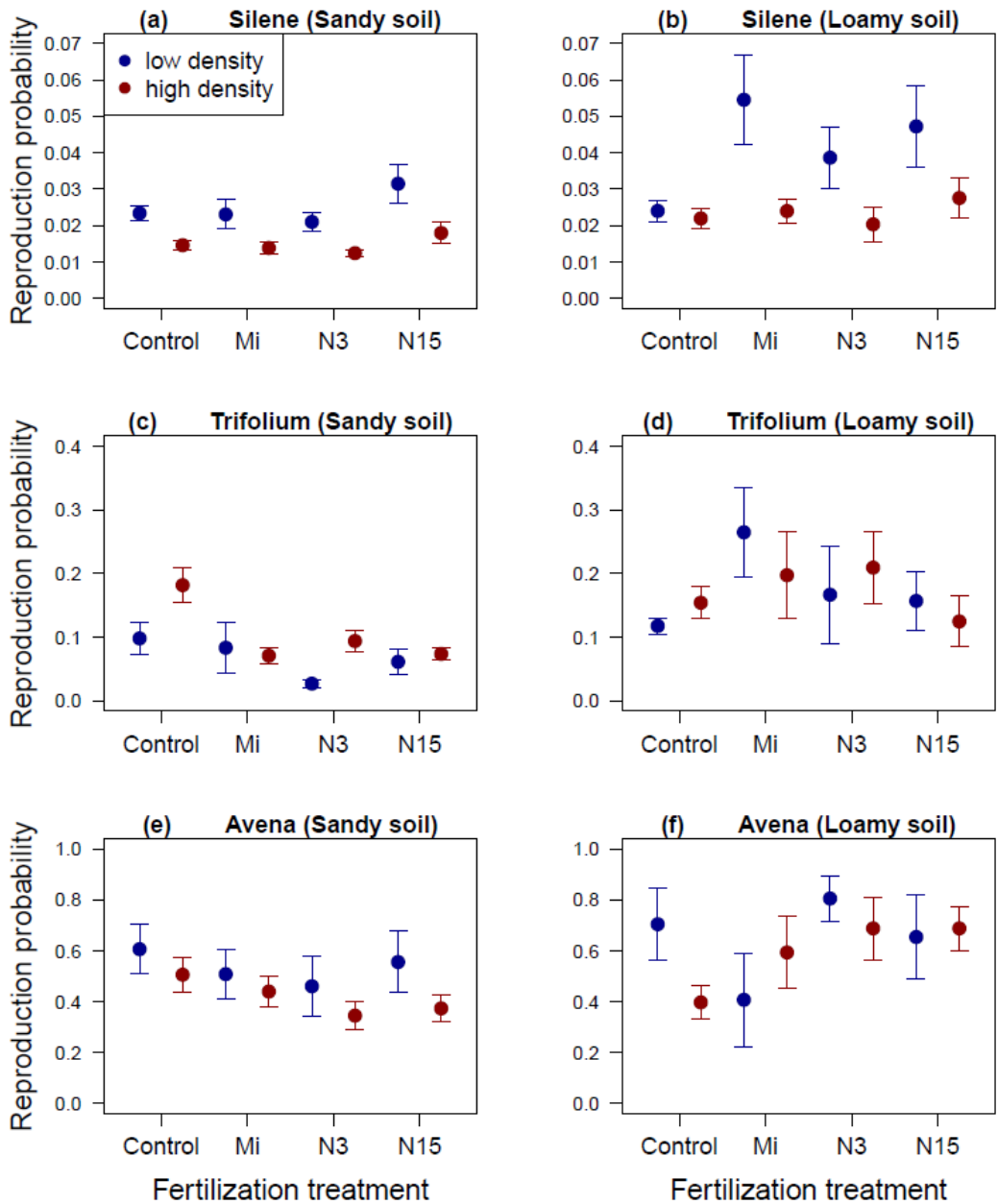

**Figure 2.** The effects of fertilization treatments and sowing density on the probability of reaching reproduction (the ratio between density in the spring and sowing density). The blue and red circles represent the means in the low and high sowing densities, respectively. The error bars are standard errors. (Mi) - addition of all nutrients (P, K, and micro) except N, (N3), and (N15) are treatments with N addition of 3g m$^{-2}$ and 15g m$^{-2}$ (together with all other nutrients). The left and right panels show the results for sandy and loamy soil types. See table S1 for statistical analyses. Note the different ranges of the y-axes.

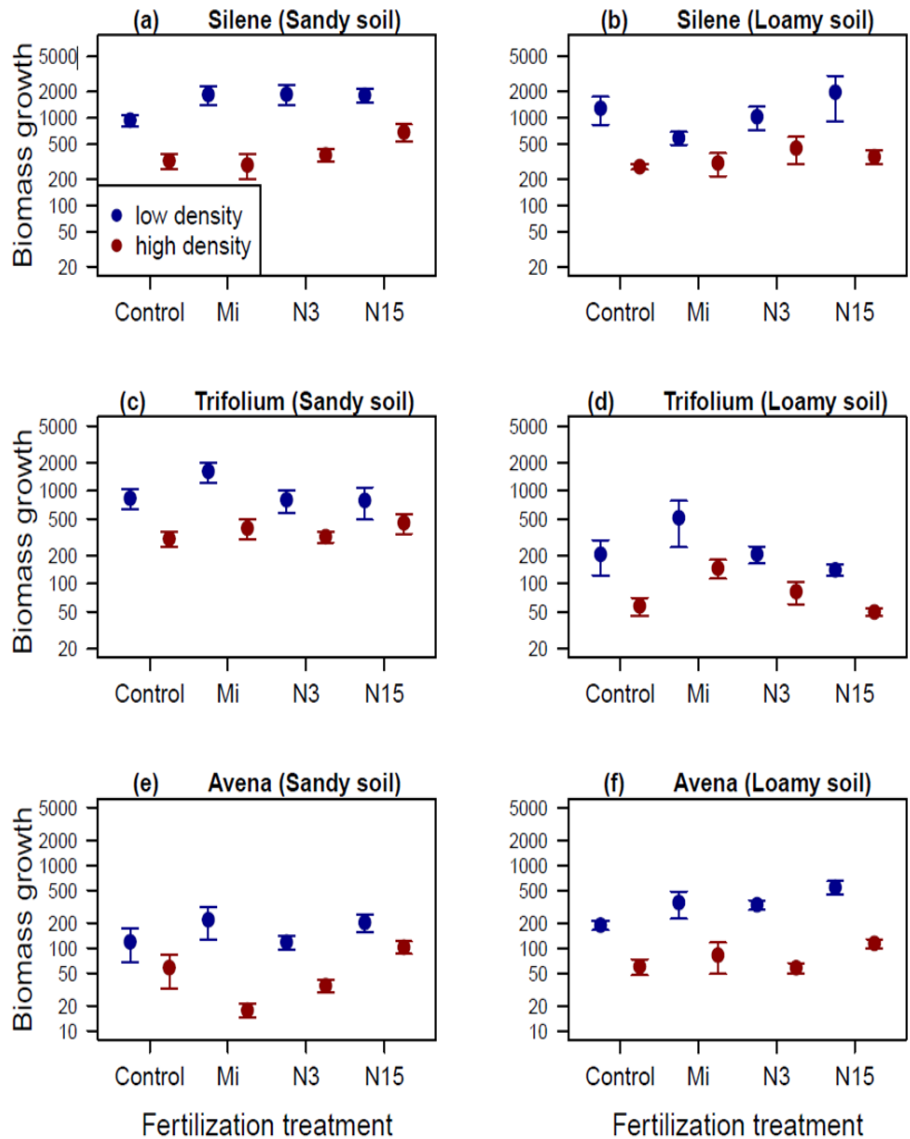

**Figure 3. The effects of fertilization treatments and sowing density on the biomass growth rate (the ratio between seed and adult masses). The blue and red circles represent the means in the low and high sowing densities. The error bars are standard errors. (Mi) - addition of all nutrients (P, K, and micro) except N, (N3), and (N15) are treatments with N addition of 3g m$^{-2}$ and 15g m$^{-2}$ (together with all other nutrients). The left and right panels show the results for sandy and loamy soil types. Note the logarithmic scale of the y-axes.**

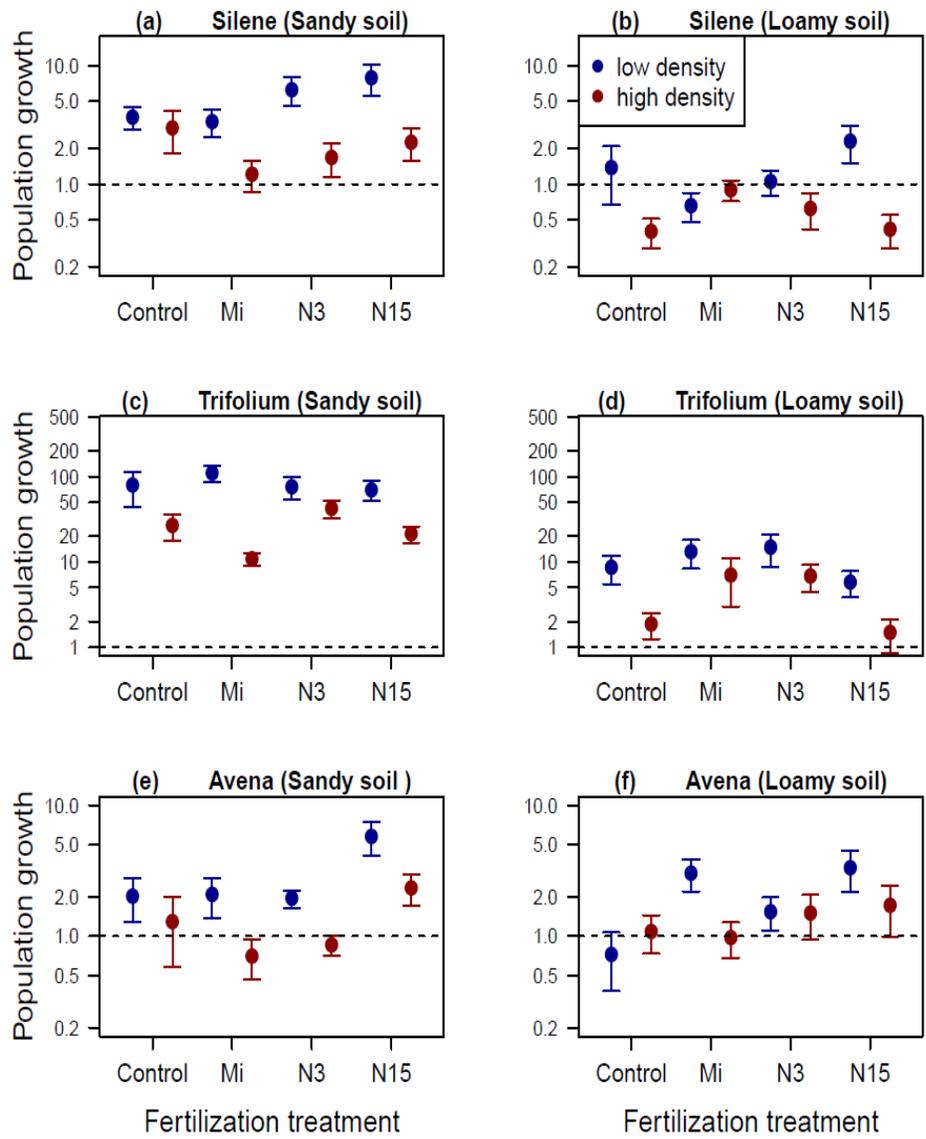

**Figure 4.** The effects of fertilization treatments and sowing density on the population growth rate (the ratio between seedlings densities in year two and year one). The dashed line represents the growth rate of one, i.e., no change in population size. The blue and red circles represent the means in the low and high sowing densities. The error bars are standard errors. (Mi) - addition of all nutrients (P, K, and micro) except N, (N3), and (N15) are treatments with N addition of 3g m$^{-2}$ and 15g m$^{-2}$ (together with all other nutrients). The left and right panels show the results for sandy and loamy soil types. Note the logarithmic scale and the different ranges of the y-axes.

## 4. DISCUSSION

We tested the role of density-dependent and density-independent mechanisms in reducing species performance with increasing N availability. In our experimental system, biomass growth and population growth decreased with sowing density for all species, implying that species interactions were predominantly negative, a necessary condition for testing N-induced density-dependent mechanisms of reduced performance. Additionally, we found that N limited growth for the forb and grass species but not for the legume species that responded positively to the addition of other resources (probably because of phosphorus limitations).

We demonstrated evidence for density-dependent and density-independent performance reductions following N addition. The legume species (*Trifolium*) declined in biomass growth under high N levels regardless of density, indicating a density-independent detrimental effect of N. In contrast, the small forb (*Silene*) experienced lower population growth with N addition but only under high density, demonstrating a density-dependent effect (competition).

Our results agree with previous empirical studies in the region (DeMalach et al., 2017b) as well as places (Xia and Wan 2008, Bobbink et al. 2010, Maskell et al. 2010, Eskelinen and Harrison 2015, DeMalach et al. 2017a, Seabloom et al. 2021, Vázquez et al. 2022), showing that tall competitive grasses like *Avena* benefit from N enrichment while legumes and forbs decline. However, our findings further suggest that this reduction is driven by different processes, density-dependent mechanisms for forbs and density-independent mechanisms for legumes. Below, we discuss the implications and interpretations of our findings.

### 4.1 Density-independent mechanisms

Many studies have shown that N enrichment affects soil chemical properties (Tian et al. 2020) and microbiomes (Lekberg et al. 2021, Huang et al. 2021) with the potential to reduce species diversity regardless of the competitive interactions. In accordance, global analyses have shown that the negative effects of N on diversity are significant even after controlling for its effects on biomass (Band et al. 2022) or light (DeMalach, 2018). Moreover, a causal link between N-induced acidification and diversity decline has been established by adding calcium to buffer soil acidification, thereby reducing species loss. However, all the above studies do not provide direct evidence for density-independent mechanisms. For example, N-induced acidification can also change competitive interactions (Yao and Feng 2022). The same reasoning applies to the soil microbiome, affecting plant performance in density-dependent and independent manners (Ke and Wan 2020). Thus, as far as we know, this is the first study to quantify the density-independent effects of N directly.

In our system, we expected density-independent mechanisms to be relatively weak because soil acidification is unlikely (high limestone content, pH ~ 7.8), and the time scale is too short for long-term changes in other soil properties (Tian et al. 2020). Nonetheless, we found a very strong density-independent response of *Trifolium*, which experienced up to a three-fold reduction in biomass growth. Furthermore, we found that Trifolium's density-independent biomass was higher in the fertile loamy soil compared with the poor sandy soil. Still, the differences between the soils could be explained by lower aeration rather than a nutrient-induced response.

A possible driver of reduced performance is N toxicity. While grass species have a rhizosheath (a layer of adhering soil particles to the root surface) acting as a "biofilm-like shield" enhancing their tolerance to N-induced stress, other species do not (Tian et al. 2022). Alternatively, the reduction in performance of *Trifolium* could be related to changes in its microbiome. Such interpretation is supported by the finding that *Trifolium*'s performance is highly reduced without its microbiome (Ayilara 2022).

**4.2 Density-dependent mechanisms**

Based on theoretical expectations, we predicted that the small forb would experience a density-dependent reduction in survival and biomass growth with increasing nitrogen availability because of asymmetric competition for light (DeMalach, Zaady, Weiner, & Kadmon, 2016). Nonetheless, we found no evidence that sensitivity to competition in reproduction probability and biomass growth increases with N availability (no interactions between density and nitrogen). Moreover, the biomass of *Silene* increased rather than declined with N regardless of density (no interactions).

A strong interaction between N availability and density was found for the population growth rate of *Silene*. While its population increased under low density, it declined under high density. We attribute the decline in population growth rate with N to a low establishment probability of this small-seeded species when there was high litter from the previous year (Thompson, 1987). This finding highlights the importance of measuring the actual population growth rate rather than using biomass growth rate or fecundity as proxies in competition experiments (Goldberg et al. 1999). Currently, empirically-calibrated models of species interactions incorporate density-dependent effects on fecundity only (Stouffer 2022), but we hope that future studies will also integrate density-dependent effects on the establishment.

**4.3 Methodological issues**

The experimental design of competition experiments along environmental gradients involves two main tradeoffs: (i) realism vs. minimizing noise and (ii) resolution of environmental conditions vs. resolution of species interactions. For the first tradeoff, we chose an intermediate setup, while for the second, we focused on maximizing the number of environmental treatments (soil types and fertilization).

The main advantage of growing plants in containers is reducing noise. We created homogenous conditions and differentiated between two soil types (the natural landscape is patchy with many intermediate levels of clay). Additionally, the containers allowed for minimizing the need to remove weeds. However, these containers cannot fully mimic the natural soil. Thus, the next challenge is testing for density-dependent and density-independent mechanisms of N enrichment in even more realistic conditions. For example, when plants grow in the natural soil and nitrogen is deposited in low quantities for a long time rather than in high quantities for a short time. Importantly, our experiment is closer to the real world than many greenhouse experiments because the plants experienced ambient climatic conditions. Furthermore, the plants grew in large containers rather than in the few-liter pots used in most competition experiments. We believe that large containers are crucial to quantify density-dependent performance because they enable niche partitioning of species varying in root depth and sufficient shading to produce light competition.

Recently, Hart, Freckleton & Levine (2018) highlighted the need to separate conspecific and heterospecific competition to calibrate a phenomenological model of community dynamics. Assuming strictly pairwise interactions and accurate estimation of the model parameters, such a model can predict long-term coexistence patterns. While we appreciate this approach's strength, it requires many competition treatments, including monocultures, all pairwise combinations, and several densities (Godoy et al. 2014). Therefore, as far as we know, it was never applied under more than two environmental conditions (Wainwright et al. 2019, Pérez-Ramos et al. 2019, Van Dyke et al. 2022). Similarly, experiments where a monoculture of each species is compared to a mixture also require many combinations. To the best of our knowledge, the monoculture-mixtures approach has not been applied to an experiment like ours that include eight types of conditions (two soil types and four nutrient treatments).

Following Goldberg and her colleagues (Goldberg & Estabrook, 1998; Rajaniemi et al., 2009), we lumped together all density-dependent effects. Of course, such pooling cannot capture all the fine details of the competition nor predict long-term coexistence. Yet, the strength of Goldberg's approach is its simplicity, i.e., it requires only two treatments, low and high density. This simplicity enabled us to investigate density-dependent and independent performance under eight environmental conditions (two soil types and four fertilization treatments). Moreover, another advantage of Goldberg's approach is that it does not require simplifying assumptions, such as interactions being always negative and strictly pairwise (Kleinhesselink et al. 2022, Stouffer 2022) or species establishment being density-independent. Indeed, despite the prevalence of negative interactions in our system, positive interactions also occur (reproduction probability of *Trifolium*). Similarly, the mounting evidence of the role of litter in the establishment of plants suggests that density-dependent effects during the establishment should not be ignored (Facelli and Pickett 1991, Foster and Gross 1998).

### 4.4 Conclusion

We showed that density-independent and density-dependent mechanisms operate together to reduce species performance under high levels of N. Specifically, we demonstrated a density-independent effect for the legume species and a density-dependent effect for the forb. It remains to be tested whether these results can be generalized to other legumes and forbs. Moreover, a critical open question is how different environmental conditions affect density-dependent and density-independent responses to N enrichment. This question is timely given the significant role of N eutrophication in driving biodiversity loss worldwide. Hence, the simplicity of the approach applied here makes it a promising way forward.

### REFERENCES

Ackerman, D. et al. 2019. Global estimates of inorganic nitrogen deposition across four decades. - Global Biogeochem. Cycles 33: 100–107.

Ayilara, I. 2022. Competition and plant-soil feedback following nitrogen addition: a greenhouse experiment.

Band, N. et al. 2022. Assessing the roles of nitrogen, biomass, and niche dimensionality as drivers of species loss in grassland communities. - Proc. Natl. Acad. Sci. U. S. A. 119: 1–11.

Bobbink, R. et al. 2010. Global assessment of nitrogen deposition effects on terrestrial plant diversity: a synthesis. - Ecol. Appl. 20: 30–59.

Borer, E. T. and Stevens, C. J. 2022. Nitrogen deposition and climate: an integrated synthesis. - Trends Ecol. Evol. xx: 1–12.

Borer, E. T. et al. 2014a. Herbivores and nutrients control grassland plant diversity via light limitation. - Nature 508: 517–520.

Borer, E. T. et al. 2014b. Finding generality in ecology: A model for globally distributed experiments. - Methods Ecol. Evol. 5: 65–73.

Britto, D. T. and Kronzucker, H. J. 2002. NH4+ toxicity in higher plants: a critical review. - J. Plant Physiol. 159: 567–584.

Chase, J. M. and Leibold, M. A. 2003. Ecological niches: linking classical and contemporary approaches. - University of Chicago Press.

Clark, C. M. et al. 2007. Environmental and plant community determinants of species loss following nitrogen enrichment. - Ecol. Lett. 10: 596–607.

Crawley, M. J. et al. 2005. Determinants of species richness in the park grass experiment. - Am. Nat. 165: 179–192.

DeMalach, N. 2018. Toward a mechanistic understanding of the effects of nitrogen and phosphorus additions on grassland diversity. - Perspect. Plant Ecol. Evol. Syst. 32: 65–72.

DeMalach, N. and Kadmon, R. 2017. Light competition explains diversity decline better than niche dimensionality. - Funct. Ecol. 31: 1834–1838.

DeMalach, N. et al. 2016. Size asymmetry of resource competition and the structure of plant communities. - J. Ecol. 104: 899–910.

DeMalach, N. et al. 2017a. Contrasting effects of water and nutrient additions on grassland communities: A global meta-analysis. - Glob. Ecol. Biogeogr. 26: 983–992.

DeMalach, N. et al. 2017b. Light asymmetry explains the effect of nutrient enrichment on grassland diversity. - Ecol. Lett. 20: 60–69.

DeMalach, N. et al. 2019. Mechanisms of seed mass variation along resource gradients. - Ecol. Lett. 22: 181–189.

Dickson, T. L. and Foster, B. L. 2011. Fertilization decreases plant biodiversity even when light is not limiting. - Ecol. Lett. 14: 380–388.

Eek, L. and Zobel, K. 2001. Structure and diversity of a species-rich grassland community, treated with additional illumination, fertilization and mowing. - Ecography (Cop.). 24: 157–164.

Eskelinen, A. and Harrison, S. 2015. Erosion of beta diversity under interacting global change impacts in a semi-arid grassland. - J. Ecol. 103: 397–407.

Eskelinen, A. et al. 2022. Light competition drives herbivore and nutrient effects on plant diversity. - Nat. 2022 6117935 611: 301–305.

Facelli, J. M. and Pickett, S. T. A. 1991. Plant litter - its dynamics and effects on plant community structure. - Bot. Rev. 57: 1–32.

Farrer, E. C. and Suding, K. N. 2016. Teasing apart plant community responses to N enrichment: the roles of resource limitation, competition and soil microbes (J Knops, Ed.). - Ecol. Lett. 19: 1287–1296.

Foster, B. L. and Gross, K. L. 1998. Species richness in a successional grassland: Effects of nitrogen enrichment and plant litter. - Ecology 79: 2593–2602.

Godoy, O. et al. 2014. Phylogenetic relatedness and the determinants of competitive outcomes (J Chave, Ed.). - Ecol. Lett. 17: 836–844.

Goldberg, D. E. and Estabrook, G. F. 1998. Separating the effects of number of individuals sampled and competition on species diversity: an experimental and analytic approach. - J. Ecol. 86: 983–988.

Goldberg, D. E. et al. 1999. Empirical approaches to quantifying interaction intensity: competition and facilitation along productivity gradients. - Ecology 80: 1118–1131.

Gough, L. et al. 2012. Incorporating clonal growth form clarifies the role of plant height in response to nitrogen addition. - Oecologia 169: 1053–1062.

Grace, J. B. et al. 2016. Integrative modelling reveals mechanisms linking productivity and plant species richness. - Nature 529: 390-+.

Gurevitch, J. et al. 1990. Competition among old field perennials at different level of soil fertility and available space. - J. Ecol. 78: 727–744.

Harpole, W. S. et al. 2016. Addition of multiple limiting resources reduces grassland diversity. - Nature 537: 93–96.

Harpole, W. S. et al. 2017. Out of the shadows: multiple nutrient limitations drive relationships among biomass, light and plant diversity. - Funct. Ecol. 31: 1839–1846.

Hart, S. P. et al. 2018. How to quantify competitive ability (H de Kroon, Ed.). - J. Ecol. 106: 1902–1909.

Hautier, Y. et al. 2009. Competition for Light Causes Plant Biodiversity Loss After Eutrophication. - Science 324: 636–638.

Huang, K. et al. 2021. Plant–soil biota interactions explain shifts in plant community composition under global change. - Funct. Ecol. 35: 2778–2788.

Hutchinson, G. E. 1957. Concluding Remarks. - Cold Spring Harb. Symp. Quant. Biol. 22: 415–427.

Isbell, F. et al. 2013. Low biodiversity state persists two decades after cessation of nutrient enrichment. - Ecol. Lett. 16: 454–460.

Ke, P. J. and Wan, J. 2020. Effects of soil microbes on plant competition: a perspective from modern coexistence theory. - Ecol. Monogr. 90: e01391.

Kleinhesselink, A. R. et al. 2022. Detecting and interpreting higher-order interactions in ecological communities. - Ecol. Lett. 25: 1604–1617.

Lamb, E. G. et al. 2009. Shoot, but not root, competition reduces community diversity in experimental mesocosms. - J. Ecol. 97: 155–163.

Lan, Z. et al. 2015. Testing the scaling effects and mechanisms of N-induced biodiversity loss: evidence from a decade-long grassland experiment. - J. Ecol. 103: 750–760.

Lekberg, Y. et al. 2021. Nitrogen and phosphorus fertilization consistently favor pathogenic over mutualistic fungi in grassland soils. - Nat. Commun. 2021 121 12: 1–8.

Maskell, L. C. et al. 2010. Nitrogen deposition causes widespread loss of species richness in British habitats. - Glob. Chang. Biol. 16: 671–679.

Midolo, G. et al. 2019. Impacts of nitrogen addition on plant species richness and abundance: A global meta-analysis. - Glob. Ecol. Biogeogr. 28: 398–413.

Miller, T. E. and Werner, P. A. 1987. Competitive Effects and Responses Between Plant Species in a First-Year Old-Field Community. - Ecology 68: 1201–1210.

Pérez-Ramos, I. M. et al. 2019. Functional traits and phenotypic plasticity modulate species coexistence across contrasting climatic conditions. - Nat. Commun. 10: 1–11.

Rajaniemi, T. K. 2003a. Evidence for size asymmetry of belowground competition. - Basic Appl. Ecol. 4: 239–247.

Rajaniemi, T. K. 2003b. Explaining productivity-diversity relationships in plants. - Oikos 101: 449–457.

Rajaniemi, T. K. et al. 2009. Community-level consequences of species interactions in an annual plant community. - J. Veg. Sci. 20: 836–846.

Sala, O. E. et al. 2000. Biodiversity - Global biodiversity scenarios for the year 2100. - Science 287: 1770–1774.

Seabloom, E. W. et al. 2021. Species loss due to nutrient addition increases with spatial scale in global grasslands. - Ecol. Lett. 24: 2100–2112.

Simkin, S. M. et al. 2016. Conditional vulnerability of plant diversity to atmospheric nitrogen deposition across the United States. - Proc. Natl. Acad. Sci. 113: 4086–4091.

Soons, M. B. et al. 2017. Nitrogen effects on plant species richness in herbaceous communities are more widespread and stronger than those of phosphorus. - Biol. Conserv. 212: 390–397.

Stevens, C. J. et al. 2004. Impact of nitrogen deposition on the species richness of grasslands. - Science 303: 1876–1879.

Stevens, C. J. et al. 2010. Nitrogen deposition threatens species richness of grasslands across Europe. - Environ. Pollut. 158: 2940–2945.

Stouffer, D. B. 2022. A critical examination of models of annual-plant population dynamics and density-dependent fecundity. - Methods Ecol. Evol. 00: 1–15.

Suding, K. N. et al. 2005. Functional- and abundance-based mechanisms explain diversity loss due to N fertilization. - Proc. Natl. Acad. Sci. U. S. A. 102: 4387–4392.

Thompson, K. 1987. Seeds and seed banks. - New Phytol. 106: 23–34.

Thompson, P. L. et al. 2020. A process-based metacommunity framework linking local and regional scale community ecology. - Ecol. Lett.: ele.13568.

Tian, Q. et al. 2020. Below-ground-mediated and phase-dependent processes drive nitrogen-evoked community changes in grasslands. - J. Ecol. 108: 1874–1887.

Tian, Q. et al. 2022. An integrated belowground trait-based understanding of nitrogen-driven plant diversity loss. - Glob. Chang. Biol. 28: 3651–3664.

Tilman, D. 1993. Species richness of experimental productivity gradients: how important is colonization limitation? - Ecology 74: 2179–2191.

Tognetti, P. M. et al. 2021. Negative effects of nitrogen override positive effects of phosphorus on grassland legumes worldwide. - Proc. Natl. Acad. Sci. U. S. A. 118: e2023718118.

Van Dyke, M. N. et al. 2022. Small rainfall changes drive substantial changes in plant coexistence. - Nat. 2022 6117936 611: 507–511.

Vázquez, E. et al. 2022. Nitrogen but not phosphorus addition affects symbiotic N2 fixation by legumes in natural and semi-natural grasslands located on four continents. - Plant Soil 2022: 1–19.

Vojtech, E. et al. 2007. Differences in Light Interception in Grass Monocultures Predict Short-Term Competitive Outcomes under Productive Conditions. - PLoS One 2: e499.

Wainwright, C. E. et al. 2019. Distinct responses of niche and fitness differences to water availability underlie variable coexistence outcomes in semi-arid annual plant communities. - J. Ecol. 107: 293–306.

Xia, J. Y. and Wan, S. Q. 2008. Global response patterns of terrestrial plant species to nitrogen addition. - New Phytol. 179: 428–439.

Yao, Q. and Feng, Y. 2022. Species existence and coexistence under nutrient enrichment in the Park Grass. - bioRxiv in press.